\documentclass[twocolumn,showpacs,preprintnumbers,amsmath,amssymb]{revtex4}
\usepackage{dcolumn}
\usepackage{bm}
\usepackage{amssymb,amsmath, amsthm,epsfig,graphicx,mathrsfs,latexsym}

\newcommand{\da}{\partial}

\begin{document}
\title{Nonlinear resonant absorption of fast magnetoacoustic waves
in strongly anisotropic and dispersive plasmas}
\author{Christopher TM Clack and Istvan Ballai}
\affiliation{Solar Physics and Space Plasma Research Centre
($SP^2RC$), Department of Applied Mathematics, University of
Sheffield, Hicks Building, Hounsfield Road, Sheffield, S3 7RH, U.K.}

\begin{abstract}
The nonlinear theory of driven magnetohydrodynamics (MHD) waves in strongly anisotropic and dispersive
plasmas, developed for slow resonance by Clack \& Ballai [Phys. Plasmas {\bf 15}(8),
2310 (2008)] and Alfv\'{e}n resonance by Clack \emph{et al.} [A\&A {\bf 494}, 317 (2009)],
is used to study the weakly nonlinear interaction of
fast magnetoacoustic (FMA) waves in a one-dimensional planar plasma.
The magnetic configuration consists of an
inhomogeneous magnetic slab sandwiched between two regions of
semi-infinite homogeneous magnetic plasmas. Laterally driven FMA
waves penetrate the inhomogeneous slab interacting with the
localized slow or Alfv\'{e}n dissipative layer and are partly
reflected, dissipated and transmitted by this region. The
nonlinearity parameter defined by Clack \& Ballai (2008) is assumed
to be small and a regular perturbation method is used to obtain
analytical solutions in the slow dissipative layer. The effect of
dispersion in the slow dissipative layer is to further decrease the
coefficient of energy absorption, compared to its standard weakly
nonlinear counterpart, and the generation of higher harmonics in the
outgoing wave in addition to the fundamental one. The absorption of external drivers
at the Alfv\'{e}n resonance is described within the linear MHD with great accuracy.
\end{abstract}

\maketitle

\vspace{0.1cm}

\section{Introduction}

The problem of interacting fast magnetoacoustic (FMA) waves with
different magnetic structures is not only important in the context
of astrophysics and solar physics, but also in laboratory plasma
devices. Space and laboratory plasmas are highly non-uniform and
dynamical systems and as a consequence they are a natural medium for
magnetohydrodynamic (MHD) waves. When the magnetic plasma
configuration is inhomogeneous in the transversal direction relative
to the ambient magnetic field a phenomenon, known as resonant
absorption, occurs (see, e.g., Appert \emph{et al.}
\citep{appert1974} and Ionson \citep{ionson1985}). Some of the wave
energy can be converted into heat in a thin layer which
embraces the ideal resonant magnetic surface when dissipative
processes are taken into account.

In the context of solar physics, the resonant coupling of waves was
first suggested by Ionson\citet{ionson1978} as a possible mechanism
for heating coronal loops. Shortly after, several studies on the
efficiency of resonant absorption in the complicated process of
coronal heating were published by, e.g., Ionson\citet{ionson1985},
Kuperus \emph{et al.}\citet{kuperus1981}, Davila\citet{davila1987} and Hollweg\citet{hollweg1990}.
The same principle was used to explain the observed loss of power of acoustic
oscillations in the vicinity of sunspots by, e.g.,
Hollweg\citet{hollweg1988}, Lou\citet{lou1990}, Sakurai \emph{et
al.}\citet{sakurai1991}, Goossens and Poedts\citet{goossens1992},
Goossens and Hollweg\citet{goossens1993} and Stenuit \emph{et
al.}\citet{stenuit1993}. All these studies dealt with the Alfv\'{e}n
resonant position. Although happening at lower frequencies, slow
resonance is also important as shown in a study by
Keppens\citet{keppens1996} where he investigated the interaction of sound
waves with hot evacuated magnetic fibrils. Most of the analytical
studies of resonant absorption were based on the linear theory due to
its relative simplicity.

A new approach to the problem of resonant absorption in the context
of high Reynolds number plasmas was given by Ruderman \emph{et
al.}\citet{ruderman1997} who developed a nonlinear theory of
resonant absorption for slow waves in isotropic plasmas. They
pointed out that nonlinearity has to be taken into account under
typical solar conditions near resonance. The theory of nonlinear
resonant slow waves was extended to strongly anisotropic plasmas in
Ballai \emph{et al.}\citet{ballai1998a} to describe conditions
typical for the solar chromosphere and corona. Over the next few years there was an
enormous amount of effort put into studying resonant absorption,
including the investigation of the effect of equilibria flows at the
slow resonance (see, e.g., Ballai and Erd\'{e}lyi
\citep{ballai1998b}), the absorption of sound waves at the slow
dissipative layers in isotropic and anisotropic plasmas (see, e.g.,
Ruderman \emph{et al.} \citep{ruderman1997b} and Ballai \emph{et
al.} \citep{ballai1998c}) and the effect of  an equilibrium flow on the
absorption of sound and FMA waves due to the coupling in the slow
continua (see, e.g., Erd\'{e}lyi and Ballai \citep{erdelyi1999a} and
Erd\'{e}lyi \emph{et al.} \citep{erdelyi2001}). In a recent paper,
Clack and Ballai\citet{clack2008} showed that in strongly
anisotropic and dispersive plasmas the dispersion, dissipation and
nonlinearity are all of the same order inside the dissipative layer.

A study by Clack \emph{et al.}\citet{clack2008b} on the nonlinear
effects at the Alfv\'{e}n dissipative layer found that nonlinearity
and dispersion are always negligible in comparison to the linear
terms describing dissipation. This implies that the linear theory is always applicable for
resonant absorption at the Alfv\'{e}n resonance if the dimensionless
amplitude of perturbations inside the dissipative layer are less
than unity. Moreover, Clack \emph{et al.}\citet{clack2008b} showed
that the largest nonlinear and dispersive terms cancel out - leaving
only small corrections to linear theory.

Many studies of resonant absorption considered only the sound (or
slow) and Alfv\'{e}n waves as excellent candidates for coronal
heating. Alfv\'{e}n waves can only carry energy along the magnetic
field lines and slow waves are only able to carry $1-2\%$ of energy
under coronal (low plasma-$\beta$) conditions. However, FMA waves
might have an important role in explaining the coronal temperatures,
as has been shown by, e.g., \u{C}ade\u{z} \emph{et al.}\citet{cadez1997}
and Cs\'{i}k \emph{et al.}\citet{csik1998}.

The aim of the present paper is to study the nonlinear (linear)
resonant interaction of externally driven FMA waves with the slow (Alfv\'{e}n)
dissipative layer in strongly anisotropic and dispersive static
plasmas. The governing equations and jump conditions derived earlier
by Clack and Ballai\citet{clack2008} and Clack \emph{et al.}\citet{clack2008b} will be used to study the
efficiency of absorption at the slow and Alfv\'{e}n resonance.
The paper is organized as follows. In the next section we
introduce the governing equations, the equilibrium state and the
fundamental assumptions which allow analytical progress. In Sec. III we
find the solutions describing the waves outside the dissipative layers.
Section IV is devoted to the nonlinear solution inside the slow
dissipative layer. In Sec. V we derive the solution inside the
Alfv\'{e}n dissipative layer. In Sec. VI we will calculate the absorption
coefficient in the case of slow/Alfv\'{e}n resonance.
Finally, in Sec. VII we summarize our results and draw our
conclusions.

\section{Governing Equations and Assumptions}

\begin{figure}[!tb]
  \centering
  \includegraphics[width=7.5cm]{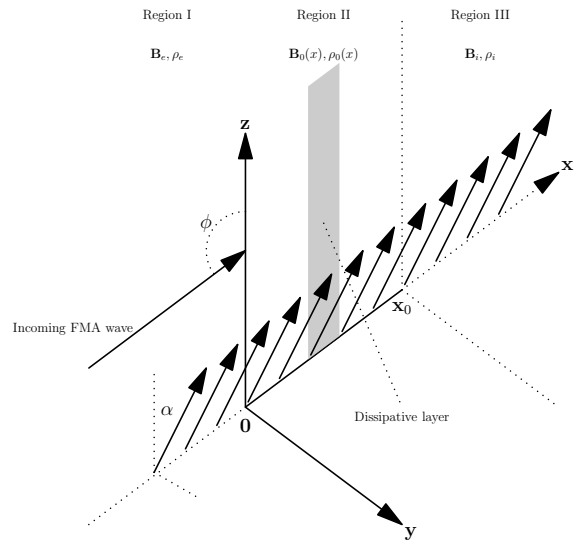}\\
  \caption{Illustration of the equilibrium state.
  Regions I ($x<0$) and III ($x>0$) contain a homogeneous magnetized
  plasma and
  Region II ($0<x<x_0$) an inhomogeneous magnetized plasma. The shaded
  strip shows the dissipative layer embracing the ideal resonant position $x_c$.} \label{fig:paperdiag1c}
\end{figure}

The dynamics and absorption of the waves will be studied in a
Cartesian coordinate system. The equilibrium state is shown in
Figure \ref{fig:paperdiag1c}. The configuration consists of an
inhomogeneous magnetized plasma $0<x<x_0$ (Region II) sandwiched
between two semi-infinite homogeneous magnetized plasmas $x<0$ and
$x>x_0$ (Regions I and III, respectively).
We have chosen this model to obtain analytical results. Our intention
is to have a model which gives us the trend in the absorption of an incident wave on a magnetic structure.
It is obvious that real magnetic structures are more complicated (and far from being fully understood), however, the magnetic field has been simplified to be unidirectional in order to make the
model more transparent, such that the role of the dispersion at the resonance and the change in the absorption can be investigated more fully, and compared to
previous studies. We took inspiration for this model from seminal studies such as Ruderman \emph{et al.}\citet{ruderman1997b}, Ballai \emph{et al.}\citet{ballai1998c}, Erd\'{e}lyi\citet{erdelyi2001}, Roberts\citet{roberts1980}, Edwin and Roberts\citet{roberts1981} and Ruderman\citet{ruderman2000}.

The equilibrium density
and pressure are denoted by $\rho$ and $p$. The equilibrium magnetic
field, $\mathbf{B}$, is unidirectional and lies in the $yz$-plane.
In what follows the subscripts $``e"$, $``0"$ and $``i"$ denote the
equilibrium quantities in the three regions (Regions I, II, III,
respectively).
It is convenient to introduce the angle, $\alpha$,
between the $z$-axis and the direction of the equilibrium magnetic
field, so that the components of the equilibrium magnetic field are;
$B_y=B\sin\alpha\mbox{ and }B_z=B\cos\alpha$.
All equilibrium quantities are continuous at the boundaries of
Region II, so they satisfy the equation of total pressure balance.
It follows from the equation of total pressure that the density ratio between Regions I and III
satisfy the relation
\begin{equation}\label{eq:densityrat}
\frac{\rho_i}{\rho_e}=\frac{2c_{Se}^2+\gamma
v_{Ae}^2}{2c_{Si}^2+\gamma v_{Ai}^2},
\end{equation}
where the squares of the Alfv\'{e}n and sound speed are
$v_{A}^2=B_0^2/\mu_0\rho_0$ and $c_{S}^2=\gamma p_0/\rho_{0}$.
Where $\mu_0$ is the magnetic permeability of free space and $\gamma$ is the adiabatic constant.
Replace the subscript $``0"$ with $``e"$ for Region I and $``i"$ for Region III.
We consider a hot magnetized plasma such that
$c_{Si}^2>c_{Se}^2,\mbox{ and }v_{Ai}^2>v_{Ae}^2$.

The objective of the present paper is to study (i) the combined
effect of nonlinearity and dispersion on the interaction of incoming
\emph{fast waves} with \emph{slow dissipative layers} and (ii) the
interaction of incoming \emph{fast waves} with \emph{Alfv\'{e}n
dissipative layers}. We, therefore, have two different criteria. For
interaction of FMA waves with the slow dissipative layer we assume that the
frequency of the incoming fast wave is within the slow continuum of
the inhomogeneous plasma, so that there is a slow resonant position at $x=x_c$ in
Region II. Interactions with the Alfv\'{e}n dissipative layer leads
to the assumption that the frequency of the incoming fast wave is
within the Alfv\'{e}n continuum of the inhomogeneous plasma,
so that there is an Alfv\'{e}n
resonant point at $x=x_a$ in Region II. This leads to the
inequality, $c_{Te}<\omega/k<c_{Ti}$, at the slow resonance. Where the square of the cusp speed, $c_T^2$, is defined by
$c_{T}^2=c_{S}^2v_{A}^2/(c_{S}^2+v_{A}^2)$. We also have the inequality, $v_{Ae}<\omega/k<v_{Ai}$,
for Alfv\'{e}n resonance.
Here $\omega$ is the frequency of the incoming fast wave and
$k=(k_x^2+k_z^2)^{1/2}$ is the wave number. Even though, in principle,
when a slow resonance occurs in this manner an Alfv\'{e}n resonance
is also present we ignore the Alfv\'{e}n resonance that occurs alongside the slow resonance as
this would complicate the analysis and obscure the results associated with the slow resonance. We study the Alfv\'{e}n resonance
separately to the slow resonance. We note that
the Alfv\'{e}n resonance would, in simple terms, act to \emph{restrict the energy} available at
the slow resonance. We intend to address the issue of coupled resonances in our next paper, where
we will show that the governing equations derived here remain the same (meaning the
work here is valid), however, the interaction of the waves between
the resonant positions changes the absorption of wave energy.

In an attempt to remove other effects from the analysis we consider
the incoming fast wave to be entirely in the $xz$-plane, i.e.
$k_y=0$. Ruderman \emph{et al.}\citet{ruderman1997b} suggests
aligning the equilibrium magnetic field with the $z$-axis, to remove
the Alfv\'{e}n resonance (if we consider planar waves) from the
analysis for slow resonance, however, this is not possible nor
necessary here. The dispersion is dependent on the angle between the
equilibrium magnetic field and the $z$-axis ($\alpha$), hence if
$\alpha=0$ the dispersion effects disappear, and we recover the
governing equation studied by Ballai \emph{et
al.}\citet{ballai1998c}.

The inequalities above guarantee that the slow
and Alfv\'{e}n resonances appears in Region II when studying in the upper
chromosphere and the solar corona, respectively. The resonant positions, therefore, are defined
mathematically as: $\omega_c=kc_T(x_c)\cos\alpha$ and $\omega_a=kv_A(x_a)\cos\alpha$.
The position of the resonant points also provides us with some information
about the plasma condition. First, in conjunction with Eq.
(\ref{eq:densityrat}) we obtain that
\begin{equation}\label{eq:rat2}
\frac{\rho_i}{\rho_e}=\frac{2c_{Se}^2+\gamma
v_{Ae}^2}{2c_{Si}^2+\gamma v_{Ai}^2}<1.
\end{equation}
Hence,  the plasma in region III is more rarefied than in Region I.
Secondly, it follows that $c_{Te}<c_{Ti}$ and the plasma in Region
III is hotter than the plasma in Region I.

The dispersion relation for the impinging propagating fast waves
takes the form
\begin{equation}\label{eq:dispersion1}
\frac{\omega^2}{k^2}=\frac{1}{2}\left\{\left(v_A^2+c_S^2\right)
+\left[\left(v_A^2+c_S^2\right)^2-4v_A^2c_S^2\cos^2\phi\right]^{1/2}\right\},
\end{equation}
where $\phi$ is the angle between the direction of propagation and
the background magnetic field within the $xz$-plane and
$\mathbf{k}=k_x\mathbf{e}_x+k_z\mathbf{e}_z$. For the sake of
simplicity, we denote $\kappa_e$ as the ratio $k_x/k_z$. Since the
equilibrium magnetic field in the $xz$-plane is aligned with the
$z$-axis, the dispersion relation (\ref{eq:dispersion1}) becomes
\begin{equation}\label{eq:dispersion2}
\frac{\omega^2}{k^2}=\frac{1}{2}
\left\{\left(v_A^2+c_S^2\right)+\left[\left(v_A^2+c_S^2\right)^2-4\frac{v_A^2c_S^2}{1+\kappa_e^2}\right]^{1/2}\right\},
\end{equation}
where $1+\kappa_e^2=1/\cos^2\phi$.

We assume the plasma is \emph{strongly} magnetized in the three
regions, such that the conditions $\omega_{i(e)}\tau_{i(e)}\gg1$ are
satisfied, here $\omega_{i(e)}$ is the ion (electron) gyrofrequency
and $\tau_{i(e)}$ is the ion (electron) collision time.
Due to the strong magnetic field, transport processes are derived from Braginskii's stress tensor (see, e.g., Braginskii\citep{braginskii1965}; Ruderman \emph{et al.}\citep{ruderman1997c}). As we deal with two separate waves (slow and Alfv\'{e}n), we will need to choose the particular dissipative process which is most efficient for these waves. For slow waves, it is a good approximation
to retain only the first term of Braginksii's expression for viscosity, namely compressional viscosity\citet{Hollweg1985}. In addition, in the solar upper atmosphere slow waves are sensitive to thermal conduction. In a strongly magnetized plasma, the thermal conductivity parallel to the magnetic field lines dwarfs the perpendicular component, hence the heat flux can be approximated by the parallel component only\citet{priest1}. On the other hand, since Alfv\'{e}n waves are transversal and incompressible they are affected by the second and third components of Braginskii's stress tensor, called shear viscosity\citet{clack2008b}. Finally, Alfv\'{e}n waves are efficiently damped by finite electrical conductivity, which becomes anisotropic under coronal conditions. The parallel and perpendicular components, however, only differ by a factor of 2, so we will only consider one of them without loss of generality. All other transport mechanisms can be neglected. For further details, please refer to, for example, Clack \emph{et al.}\citet{clack2008b}, Braginskii\citet{braginskii1965}, Ruderman \emph{et al.}\citet{ruderman1997c}, Hollweg\citet{Hollweg1985}, Priest\citet{priest1} and Porter \emph{et al.}\citet{porter1994}.

The dynamics of nonlinear resonant MHD waves in anisotropic and
dispersive plasmas was studied by Clack and Ballai\citet{clack2008}
and Clack \emph{et al.}\citet{clack2008b}. They derived the
governing equations and connection formulae necessary to study
resonant absorption in slow/Alfv\'{e}n dissipative layers. We recall the key steps and necessary
results found by Clack and Ballai\citet{clack2008} and Clack
\emph{et al.}\citet{clack2008b}.

The efficiency of dissipation, when studying slow dissipative
layers, in an anisotropic plasma is given by the (compressional)
viscous Reynolds number ($R_{e(c)}$) and the Pechlet number ($P_e$),
combining to define the total Reynolds number: ${R^{-1}_c}={R^{-1}_{e(c)}}+{P^{-1}_e}$,
where $R_{e(c)}$ and $P_e$ are defined by
$R_{e(c)}=v_{ch}l_{ch}\rho_{0_c}/\bar{\eta}_{0}$ and
$P_e=v_{ch}l_{ch}\rho_{0c}\widetilde{R}/\bar{\kappa}_{\parallel}$.
Here $v_{ch}$ is the characteristic velocity (e.g. the slow
magnetoacoustic velocity at $x=x_c$), $l_{ch}$ is the characteristic
length, $\rho_{0c}=\rho_0(x_c)$, $\widetilde{R}$ denotes the gas
constant and $\kappa_{\parallel}$ is the coefficient of thermal
conductivity parallel to the equilibrium magnetic field lines. The
efficiency of dissipation, when studying Alfv\'{e}n dissipative
layers, in an anisotropic plasma is measured in a slightly different way. Now
dissipative processes are described by the (shear) viscous Reynolds number ($R_{e}$) and the magnetic
Reynolds number ($R_m$), combining to define the total Reynolds
number: ${R^{-1}_a}={R^{-1}_{e}}+{R^{-1}_m}$, where
$R_{e}=v_{ch}l_{ch}\rho_{0_a}/\bar{\eta}_{1}$ and $R_m=v_{ch}l_{ch}/\bar{\eta}$.
Here $v_{ch}$ is the characteristic velocity (e.g. the Alfv\'{e}n
velocity at $x=x_a$), $l_{ch}$ is the characteristic length,
$\eta_1$ is the coefficient of shear viscosity and $\eta$ is the
coefficient of finite electrical resistivity.
Originally, these total Reynolds numbers were introduced based on intuition, simplicity and linear theory (see, e.g., Sakurai \emph{et al.} \citep{sakurai1991}, Goossens \emph{et al.} \citep{goossens1994} and Goossens and Ruderman \citep{goossens1995}). However, it turned out that using these definitions the \emph{strength} of dissipation is the same order of magnitude as the inverse of the total Reynolds numbers.
Under chromospheric
and coronal conditions $R\gg1$ which means that dissipation is only
important inside the dissipative layer. Far away from the
dissipative layer amplitudes are small, therefore we can use the
linear ideal MHD equations to describe the plasma motions far from
the resonant position. These equations can be reduced to a system of
coupled first order PDE's for the total pressure perturbation, $P$,
and the normal component of the velocity, $u$,
\begin{equation}\label{eq:linearmhdeq}
\frac{\da u}{\da x}=\frac{V}{F}\frac{\da P}{\da\theta},\quad \frac{\da P}{\da x}=\frac{\rho_0 A}{V}\frac{\da u}{\da\theta}.
\end{equation}
Here
\begin{align}\label{eq:FAC}
F&=\frac{\rho_0 A C}{V^4-V^2\left(v_A^2+c_S^2\right)+v_A^2
c_S^2\cos^{2}\alpha},\nonumber\\
C&=\left(v_A^2+c_S^2\right)\left(V^2-c_T^2\cos^{2}\alpha\right),\nonumber\\
A&=V^2-v_A^2\cos^{2}\alpha.
\end{align}
The system (\ref{eq:linearmhdeq}) describes the wave motion far from
the ideal resonant position. The singularities in the coefficients
$A$ and $C$ give the conditions of Alfv\'{e}n and slow resonance.
All perturbations depend on the combination $\theta=z-Vt$, where
$V=\omega/k$ is the phase speed.

Inside the thin dissipative layers (where the dynamics is described by the nonlinear and dissipative MHD
equations) embracing the ideal resonant
surfaces ($x=x_c$, $x=x_a$) we must use the governing equations
derived by Clack and Ballai\citet{clack2008} and Clack \emph{et
al.}\citet{clack2008b}. The characteristic thickness of the slow
dissipative layer, $\delta_c$, is
\begin{equation}\label{eq:disslayer}
\delta_c=\frac{V^3k\lambda}{|\Delta_c|\left(c_{Sc}^2+v_{Ac}^2\right).}
\end{equation}
Here $k=2\pi/L$ with $L$ the wavelength, the subscript ``c"
indicates that the quantity has been calculated at the slow resonant
position. The quantity $\lambda$ is defined by
\begin{equation}\label{eq:lambda}
\lambda=\frac{\eta_0\left(2v_{A_c}^2+3c_{S_c}^2\right)^2}
{3\rho_{0_c} v_{A_c}^2 c_{S_c}^2}+\frac{(\gamma-1)^2
\kappa_{\parallel}\left(v_{A_c}^2+c_{S_c}^2\right)}{\gamma\rho_{0_c}\widetilde{R}c_{S_c}^2},
\end{equation}
and $\Delta_c$ is simply the gradient of the cusp speed given by
$\Delta_c=-\left(dc_T^2/dx\right)_c\cos^2\alpha$.
Clack and Ballai\citet{clack2008} showed that nonlinearity and
dispersion are important in the slow dissipative layer if the
nonlinearity parameter is greater than unity, $N^2=\epsilon
R_c^2\gtrsim 1$, where $\epsilon$ is the dimensionless wave amplitude
far from the dissipative layer.
The concept of nonlinear parameters was introduced by Ruderman \emph{et al.}\citet{ruderman1997}
for slow waves and Clack \emph{et al.}\citet{clack2008b} for Alfv\'{e}n waves. The two parameters are different not
only in their form but also in the values the Reynolds numbers take. In the case of
slow waves (damped by compressional viscosity, i.e. the first term in the
Braginskii's viscosity tensor) the Reynolds number that corresponds to a
characteristic length of $200{\rm Mm}$, a speed of $200{\rm kms^{-1}}$, a density of $10^{-13}{\rm kgm^{-3}}$
and a compressional viscosity coefficient of $5\times10^{-2}{\rm kgm^{-1}s^{-1}}$ is about
$80$. Alfv\'{e}n waves are efficiently damped by shear viscosity which is given by the
second and third coefficients of the Bragisnkii's tensor (here denoted
cumulatively as $\eta_1$). Since $\eta_1=\eta_0/(\omega_i\tau_i)^2$ and under
coronal conditions $\omega_i\tau_i$ is of the order of $10^5$, we obtain that the
coefficient of shear viscosity is about $10$ orders of magnitude smaller than the
coefficient of compressional viscosity. Now, using the characteristic speed of
$1000{\rm kms^{-1}}$, the Reynolds number used in calculating the nonlinear parameter in
the case of Alfv\'{e}n nonlinearity is $4\times 10^{12}$. The nonlinearity parameter for resonant Alfv\'{e}n waves is $\epsilon R^{2/3}\ll1$. However, it was shown by Clack \emph{et al.}\citet{clack2008b} that the waves in this situation remain linear anyway. In the present paper, therefore, we do not need the nonlinearity parameter for resonant Alfv\'{e}n waves. The characteristic thickness of the Alfv\'{e}n dissipative
layer, $\delta_a$, is
\begin{equation}\label{eq:disslayeralf}
\delta_a=\left[\frac{V}{k|\Delta_a|}\left(\bar{\eta}+\frac{\bar{\eta_{1}}}{\rho_{0a}}\right)\right]^{1/3},
\end{equation}
with $\Delta_a$ being the gradient of the Alfv\'{e}n speed
given by $\Delta_a=-\left(dv_A^2/dx\right)_a\cos^2\alpha$.

The governing equation inside the slow dissipative layer
is\citet{clack2008}
\begin{multline}\label{eq:governing2}
\sigma_c\frac{\da q_c}{\da\theta}+\Lambda q_c\frac{\da
q_c}{\da\theta}
-k^{-1}\frac{\da^2 q_c}{\da\theta^2}\\
-\Psi\frac{\da q_c}{\da\sigma}\frac{\da q_c}{\da\theta}=
-\frac{kV^4}{\rho_{0_c}v_{A_c}^2|\Delta_c|}\frac{d\widetilde{P}}{d\theta},
\end{multline}
where
\begin{align}
&\sigma_c=\frac{x-x_c}{\delta_c},\\
&\Lambda=R^2\frac{v_{A_c}^4|\Delta|\left[(\gamma+1)v_{A_c}^2+3c_{S_c}^2\right]}{kV^8},\label{eq:Lambda}\\
&\Psi=R^2\frac{\chi|\Delta|^2 c_{S_c}^2 v_{A_c}^2
(v_{A_c}^2+c_{S_c}^2)\sin\alpha}{kV^{13}},\label{eq:psi}\\
&\chi=\eta\omega_e\tau_e.
\end{align}
Here the first term of the governing equation appears due to the inhomogeneity in the cusp
speed, the second term describes the nonlinearity of waves, the
third term stands for the dissipative effects while the last term on
the left-hand side describes the nonlinear dispersive effects
generated after taking into account Hall currents by Clack and
Ballai\citet{clack2008}. The term on the right-hand side can be
considered as a driver. We also note that $q_c(\sigma_c,\theta)$ is
the dimensionless component of velocity parallel to the equilibrium
magnetic field and $\chi=\eta\omega_e\tau_e$ is the coefficient of
Hall conduction\citet{clack2008}.

The governing equation inside the Alfv\'{e}n dissipative layer
is\citet{clack2008b}
\begin{equation}\label{eq:governingalf}
\sigma_a\frac{\da q_a}{\da\theta}-k\frac{\da^2
q_a}{\da\sigma_a^2}=\frac{k\sin\alpha}{\rho_{0_a}|\Delta_a|}\frac{d\widetilde{P}}{d\theta},
\end{equation}
with
$\sigma_a=(x-x_a)/\delta_a$.
Here $q_a(\sigma_a,\theta)$ is the dimensionless
component of velocity perpendicular to the equilibrium magnetic
field. We should point out here that although nonlinearity and dispersion
have been considered when deriving the dynamics of the Alfv\'{e}n
resonance, the governing equation remains linear regardless of the
degree of nonlinearity (for details see Clack \emph{et al.} \citep{clack2008b}).

When studying resonant MHD waves, we are generally not interested in
the solution inside the dissipative layer and can consider the
dissipative layer as a surface of discontinuity. Instead, we solve the system
(\ref{eq:linearmhdeq}) and match the solutions at the boundaries of
the discontinuity using connection formulae. These connection
formulae determine the jumps in $u$ and $P$ across the dissipative
layer. In the context of solar plasmas, they were first introduced by Sakurai \emph{et
al.}\citet{sakurai1991}. It was shown by Clack and
Ballai\citet{clack2008} (in complete agreement with Ruderman
\emph{et al.}\citet{ruderman1997} and Ballai \emph{et
al.}\citet{ballai1998a}) that the first connection formula is
$\left[P\right]=0$,
where the square brackets denote the jump across the dissipative
layer. It can also be shown, in a similar manner, that the same jump
condition exists for the Alfv\'{e}n resonance. The second connection
formula for slow resonance can only be written in implicit form,
i.e.
\begin{equation}\label{eq:ujump}
\left[u_c\right]=-\frac{V}{k\cos^2\alpha}\mathscr{P}\int_{-\infty}^{\infty}\frac{\da
q_c}{\da\theta}\mbox{ }d\sigma,
\end{equation}
where we use the Cauchy principal value of the integral because the
integral is divergent at infinity. As a result we must solve Eqs.
(\ref{eq:linearmhdeq}) and (\ref{eq:governing2}) along with the
boundary conditions, $\left[P\right]=0$ and Eq. (\ref{eq:ujump}). In an
attempt to follow the same procedure utilized for finding solutions
at the slow resonance we can write the jump in the normal component
of velocity for the Alfv\'{e}n resonance in an implicit form.
For the sake of brevity, we do not show the derivation here, but it
follows the procedure to find the jump in the normal component of
velocity completed by Clack and Ballai\citet{clack2008}. This jump
is given by
\begin{equation}\label{eq:ujumpalf}
\left[u_a\right]=\frac{V\sin\alpha}{k}\mathscr{P}\int_{-\infty}^{\infty}\frac{\da
q_a}{\da\theta}\mbox{ }d\sigma.
\end{equation}

Finally, we should note some critical assumption we make to allow
analytical progress. From the very beginning we must assume that the
nonlinearity parameter is small so that regular perturbation theory
can be applied at the slow resonance. We also assume that the inhomogeneous region is thin
in comparison with the wavelength of the impinging wave, i.e.
$kx_0\ll1$. Ruderman\citet{ruderman2000} investigated the absorption
of sound waves at the slow dissipative layer in the limit of strong
nonlinearity. In his analysis nonlinearity dominated dissipation
in the \emph{resonant layer} which embraces the dissipative layer.
He concluded that nonlinearity decreases absorption in the long
wavelength approximation, but increases it at intermediate values of $kx_0$,
however, the increase is never more than $20\%$. To the best of our
knowledge, at present, we cannot solve the governing equation
(\ref{eq:governing2}) in the limit of strong nonlinearity due to the
nonlinear dispersive term, therefore we restrict our analysis to the
weak nonlinear limit. We mention that no such assumptions are needed
for studying the Alfv\'{e}n dissipative layer since the governing
equation (\ref{eq:governingalf}) is linear.

\section{Solutions Outside the Dissipative Layers}

In what follows we derive a solution for the system
(\ref{eq:linearmhdeq}) in Regions I, II and III. In Region II we
only find the solution outside the dissipative layers. Section
IV is devoted to finding a solution to Eq. (\ref{eq:governing2})
inside the slow dissipative layer and Section V is used to find a
solution to Eq. (\ref{eq:governingalf}) inside the Alfv\'{e}n
dissipative layer. Outside the dissipative layers, the solutions take
identical forms.

\subsection{Region I}

The solution of Eq. (\ref{eq:linearmhdeq}) in Region I is given
in the form of an incoming and outgoing fast wave of the form
\begin{align}
&P=\epsilon \left\{p_e\cos\left[k\left(\theta+\kappa_e x\right)\right]+A\cos\left[k\left(\theta-\kappa_e x\right)\right]\right\},\label{eq:inoutfastwave1}\\
&u=\epsilon\frac{\kappa_eV\left\{p_e\cos\left[k\left(\theta+\kappa_e x\right)\right]-A\cos\left[k\left(\theta-\kappa_e x\right)\right]\right\}}
{\rho_e\left(V^2-v_{Ae}^2\cos^2\alpha\right)},\label{eq:inoutfastwave2}
\end{align}
where $\epsilon\ll1$ is the dimensionless amplitude of perturbation
far from the dissipative layer. The frequency of the incoming wave
is given by Eq. (\ref{eq:dispersion2}) and must lie within the slow or Alfv\'{e}n continuum depending on which dissipative layer
we are studying. The first term in Eqs. (\ref{eq:inoutfastwave1})
and (\ref{eq:inoutfastwave2}) describes the incoming wave, while the
second term describes the outgoing wave which will be obtained in
Section IV for slow dissipative layers and in Section V for
Alfv\'{e}n dissipative layers.

\subsection{Region II}

In Region II, the equation for the total pressure, $P$, is obtained by eliminating
$u$ from the system (\ref{eq:linearmhdeq}),
\begin{equation}\label{eq:pressure1}
F\frac{\da}{\da x}\left[\frac{1}{\rho_0\left(V^2-v_A^2\cos^2\alpha\right)}\frac{\da P}{\da x}\right]=\frac{\da^2P}{\da\theta^2}.
\end{equation}
Since we have assumed $kx_0\ll1$, the ratio of the right-hand side
and the left-hand side is of the order of $k^2x_0^2$. It follows that
\begin{equation}\label{eq:pressure2}
\frac{\da P}{\da x}=\rho_0\left(V^2-v_A^2\cos^2\alpha\right)f(\theta)+\mathscr{O}(k^2x_0^2),
\end{equation}
where the function $f(\theta)$ is determined by the second equation of (\ref{eq:linearmhdeq})
and the boundary conditions at $x=0$. Equation (\ref{eq:pressure2}) yields
\begin{equation}\label{eq:pressure3}
P=\widetilde{P}(\theta)+f(\theta)\int_{0}^{x}\rho_0\left[V^2-v_A^2\cos^2\alpha\right]dx+\mathscr{O}(k^2x_0^2).
\end{equation}
The function $\widetilde{P}(\theta)$ has to be determined by the boundary
conditions at $x=0$. It can be shown that, because $\left[P\right]=0$,
the functions $f(\theta)$ and $\widetilde{P}(\theta)$ take the same values throughout Region II.
Noting that the second term in Eq. (\ref{eq:pressure3}) is of the order of $kx_0$ we
can express $P$ in a simplified form
\begin{equation}\label{eq:pressure4}
P=\widetilde{P}(\theta)+\left(kx_0\right)P'(x,\theta)+\mathscr{O}(k^2x_0^2).
\end{equation}

\subsection{Region III}

To derive the governing equation for Region III we eliminate the normal
component of the velocity from the system (\ref{eq:linearmhdeq}) to arrive at
\begin{equation}\label{eq:governingIII}
\frac{\da^2 P}{\da x^2}+\kappa_i^2\frac{\da^2 P}{\da\theta^2}=0,
\end{equation}
where $\kappa_i^2$ is defined as
\begin{equation}\label{eq:kappai}
\kappa_i^2=-\frac{V^4-V^2\left(c_{Si}^2+v_{Ai}^2\right)+c_{Si}^2v_{Ai}^2\cos^2\alpha}
{\left(c_{Si}^2+v_{Ai}^2\right)\left(V^2-c_{Ti}^2\cos^2\alpha\right)}.
\end{equation}
Since, for slow dissipative layers, $V<c_{Ti}\cos\alpha$, it follows
that $\kappa_i^2>0$. It also follows that for Alfv\'{e}n dissipative
layers $\kappa_i^2>0$ because $V>c_{Ti}\cos\alpha>v_{Ae}\cos\alpha$.
Therefore, Eq. (\ref{eq:governingIII}) is an elliptical differential
equation and the wave motion is evanescent in Region III.
In reality, there could be wave leakage. The existence of wave leakage depends on the
profile of the slow and Alfv\'{e}n speeds in the inhomogeneous region (Region II). For simplicity,
we have assumed that the slow and Alfv\'{e}n resonances take place at a single location (obviously different for the two resonances),
which means the profiles of the slow and Alfv\'{e}n speeds are monotonically increasing inside Region II. Should we have a more complex model,
the possibility of wave leakage would need to be taken into account.

\section{Weak Nonlinear Solution Inside the Slow Dissipative Layer}

Since we are not able to solve the governing equation
(\ref{eq:governing2}) inside the slow dissipative layer
analytically, we consider the limit of weak nonlinearity
($N^2\ll1$). In accordance with this assumption we rewrite the
governing equation (\ref{eq:governing2}) and the jump condition
(\ref{eq:ujump}) as
\begin{multline}\label{eq:governingproper}
\sigma\frac{\da\overline{q_c}}{\da\theta}+\epsilon^{-1}\zeta\left(\frac{\Lambda}{\Psi}\right)
\overline{q_c}\frac{\da\overline{q_c}}{\da\theta}-\epsilon^{-1}\zeta\frac{\da\overline{q_c}}{\da\sigma}
\frac{\da\overline{q_c}}{\da\theta}\\
-k^{-1}\frac{\da^2\overline{q_c}}{\da\theta^2}
=-\frac{V^4}{\rho_{0c}v_{Ac}^4|\Delta_c|x_0}\frac{dP_c}{d\theta},
\end{multline}
\begin{equation}\label{eq:ujump1}
\left[u_c\right]=-\frac{Vx_0}{\cos^2\alpha}\mathscr{P}\int_{-\infty}^{\infty}\frac{\da
\overline{q_c}}{\da\theta}\mbox{ }d\sigma,
\end{equation}
where
\begin{equation}\label{eq:dimensionlesspaprameter}
\overline{q_c}=\frac{q_c}{kx_0},\quad
\zeta=\frac{kx_0D_d^2\Psi}{R^4},\quad D_d^2=\epsilon R^4=R^2N^2.
\end{equation}
Note that $\zeta$ is of the order of $\epsilon R^2$, the ratio
$(\Lambda/\Psi)$ is of the order of unity and $\overline{q_c}$ is of
the order of $\epsilon$. In what follows we drop the bar notation
and for the rest of this section we drop the subscript ``c" on the
dimensionless variable $q$.

We proceed by using a regular perturbation method and look for
solutions in the form
\begin{equation}\label{eq:expansions}
f=\epsilon\sum_{n=1}^{\infty}\zeta^{n-1}f_n,
\end{equation}
where $f$ represents any of the quantities $P$, $u$ and $q$.

\subsection{First order approximation}

In the first order approximation, from Eq. (\ref{eq:governingproper}), we obtain
\begin{equation}\label{eq:firstordereq}
\sigma\frac{\da q_1}{\da\theta}-k^{-1}\frac{\da^2q_1}{\da\theta^2}=-\frac{V^4}{\rho_{0c}v_{Ac}^4|\Delta|x_0}\frac{dP_{1c}}{d\theta}.
\end{equation}
Since the total pressure, $P$, is continuous throughout the dissipative layer
\emph{and} is periodical with respect to $\theta$, we look for a solution
in the form
$g_1=\Re(\hat{g}_1e^{ik\theta})$,
where $g_1$ represents $P_1$, $u_1$ and $q_1$ and $\Re$ indicates the
real part of a quantity.

In Region I the solutions for the pressure and velocity
exactly recover the results found in linear theory, i.e.
\begin{align}
&\hat{P}_1=p_ee^{ik\kappa_ex}+A_1e^{-ik\kappa_ex},\label{eq:pressureeq1}\\
&\hat{u}_1=\frac{\kappa_eV\left(p_ee^{ik\kappa_ex}-A_1e^{-ik\kappa_ex}\right)}
{\rho_e\left(V^2-v_{Ae}^2\cos^2\alpha\right)},\label{eq:ueq1}
\end{align}
where $A_1$ (and subsequent values of $A_i$) is the amplitude of the outgoing wave.
The first terms of the right-hand side of $\hat{P}_1$ and $\hat{u}_1$ represent
the incoming wave, while the second terms are the outgoing (reflected) wave.
The continuity of the total pressure perturbation at $x=0$ and $x=x_0$ in combination
with Eq. (\ref{eq:pressure4}) yields $\hat{P}_1$, in Region II, as
\begin{equation}\label{eq:pressureeq2}
\hat{P}_1=p_e+A_1+(kx_0)\hat{h}_1,
\end{equation}
where $\hat{h}_n=\hat{h}_n(x)=\hat{P}_n'(x)-\hat{P}_n'(0),\quad n\geq1$.
The solution in Region III is obtained by using Eqs. (\ref{eq:linearmhdeq}),
(\ref{eq:governingIII}) and (\ref{eq:pressureeq2}) with the continuity
conditions at $x=x_0$. The solution takes the form
\begin{align}
&\hat{P}_1=\left\{p_e+A_1+(kx_0)\hat{h}_1\right\}e^{-k\kappa_i(x-x_0)},\label{eq:pressureeq3}\\
&\hat{u}_1=\frac{i\kappa_iV\left\{p_e+A_1+(kx_0)\hat{h}_1\right\}}
{\rho_i\left(V^2-v_{Ai}^2\cos^2\alpha\right)}e^{-k\kappa_i(x-x_0)}.\label{eq:ueq3}
\end{align}
Utilizing the fact that $\hat{u}_1$ is continuous at $x=0$ and $x=x_0$,
and employing Eqs. (\ref{eq:linearmhdeq}) and (\ref{eq:pressureeq2}) we
find that the jump in the normal component of velocity across the dissipative layer is
\begin{multline}\label{eq:jumpinu11}
\left[\hat{u}_1\right]=\frac{i\kappa_iV(p_e+A_1)}{\rho_i\left(V^2-v_{Ai}^2\cos^2\alpha\right)}
-\frac{\kappa_eV(p_e-A_1)}{\rho_e\left(V^2-v_{Ae}^2\cos^2\alpha\right)}\\
-ikV\left(p_e+A_1\right)\mathscr{P}\int_{0}^{x_0}F^{-1}(x)\mbox{ }dx\\
-ikV(kx_0)\mathscr{P}\int_{0}^{x_0}\frac{\hat{h}_1(x)}{F(x)}\mbox{
}dx,
\end{multline}
where the expression of $F(x)$ is given by Eq. (\ref{eq:FAC}).

Solving Eq. (\ref{eq:firstordereq}) reveals $\hat{q}_1$ to be
\begin{equation}\label{eq:q1}
\hat{q}_1=-\frac{V^4\left(p_e+A_1\right)\left\{1+\mathscr{O}(kx_0)\right\}}
{\rho_{0c}v_{A_c}^2|\Delta|x_0\left(\sigma-i\right)}.
\end{equation}
Substitution of this result into Eq. (\ref{eq:ujump1}) leads to
another definition of the jump in the normal component of velocity
across the dissipative layer, namely,
\begin{equation}\label{eq:jumpinu12}
\left[\hat{u}_1\right]=\frac{-\pi kV^5\left(p_e+A_1\right)\left\{1+\mathscr{O}(kx_0)\right\}}
{\rho_{0c}v_{Ac}^4|\Delta|\cos^2\alpha}.
\end{equation}
Comparing Eqs. (\ref{eq:jumpinu11}) and (\ref{eq:jumpinu12}) we
obtain that
\begin{equation}\label{eq:A1}
A_1=-p_e\frac{\tau-\mu+i\upsilon}{\tau+\mu+i\upsilon}+\mathscr{O}(k^2x_0^2),
\end{equation}
where
\begin{align}
\tau &=\frac{\pi kV^5}{\rho_{0c}v_{Ac}^4|\Delta|\cos^2\alpha},\quad
\mu =\frac{\kappa_eV}{\rho_e\left(V^2-v_{Ae}^2\cos^2\alpha\right)}\nonumber\\
\upsilon
&=\frac{\kappa_iV}{\rho_i\left(V^2-v_{Ai}^2\cos^2\alpha\right)}
-kV\mathscr{P}\int_{0}^{x_0}F^{-1}(x)\mbox{
}dx.\label{eq:taumuupsilon}
\end{align}
When deriving Eq. (\ref{eq:A1}) we have employed the estimate that
$k\mathscr{P}\int_{0}^{x_0}\hat{h}_n(x)/F(x)\mbox{ }dx=\mathscr{O}(kx_0)$.
The quantity $A_1$ is a complex value. This means that the outgoing (reflected) wave
has a phase alteration compared with the incoming wave. The \emph{true amplitude} of the outgoing wave is given by $\widetilde{A_1}=(A^{2}_{1(r)}+A^{2}_{1(im)})^{1/2}$ (where the subscripts $``r"$ and $``im"$ mean the real and imaginary parts, respectively). The Fourier analysis allows $A_1$ to be complex. In general, a complex value of $A_n$ means the true amplitude of the outgoing harmonic is defined as above and a phase of the outgoing wave is shifted by $\tan^{-1}(A^{2}_{n(im)}/A^{2}_{n(r)})$. This definition of $A_n$ applies to all subsequent orders of approximation.

In Ruderman \emph{et al.} \cite{ruderman1997b} and Ballai \emph{et
al.} \cite{ballai1998c} a similar procedure was carried out. Our
results are similar with theirs if we consider $B_e=0$ and
$\alpha=0$. This conclusion is not surprising because the first
order approximation with respect to the nonlinearity parameter
coincides with linear theory. In addition, dispersion due to the
Hall effect at the slow resonance does not alter linear theory
either since dispersion effects appear as a nonlinear term in the
governing equation.

\subsection{Second order approximation}

Nonlinear effects start to be important from the second order approximation onwards,
but they are always due to the nonlinear combination of lower order harmonics. In this
order of approximation Eq. (\ref{eq:governingproper}) is reduced to
\begin{multline}\label{eq:secondordereq}
\sigma\frac{\da q_2}{\da\theta}-k^{-1}\frac{\da^2q_2}{\da\theta^2}
=-\frac{V^4}{\rho_{0c}v_{Ac}^4|\Delta|x_0}\frac{dP_{2c}}{d\theta}\\
-q_1\frac{\da q_1}{\da\theta}+\frac{\da q_1}{\da\sigma}\frac{\da q_1}{\da\theta}.
\end{multline}
Taking advantage of the form of the first order approximation terms enables us to
rewrite the second term on the right-hand side of this equation as
\begin{equation}\label{2ndform1}
q_1\frac{\da q_1}{\da\theta}=\Re\left(\frac{ik}{2}\hat{q}_1^2e^{2ik\theta}\right).
\end{equation}
Since the nonlinear terms are proportional to $\Re\left(e^{2ik\theta}\right)$ it
is appropriate to seek a solution of the form
$g_2=\Re\left(\hat{g}_2e^{2ik\theta}\right)$,
where $g_2$ represents $P_2$, $u_2$ and $q_2$.

Using the same techniques as in the first order approximation,
it is straightforward to find the jump in the normal component of velocity in Region II
\begin{multline}\label{eq:jumpinu21}
\left[\hat{u}_2\right]=\frac{i\kappa_iVA_2}{\rho_i\left(V^2-v_{Ai}^2\cos^2\alpha\right)}
+\frac{\kappa_eVA_2}{\rho_e\left(V^2-v_{Ae}^2\cos^2\alpha\right)}\\
-2ikVA_2\mathscr{P}\int_{0}^{x_0}F^{-1}(x)\mbox{ }dx\\
-2ikV(kx_0)\mathscr{P}\int_{0}^{x_0}\frac{\hat{h}_2(x)}{F(x)}\mbox{ }dx.
\end{multline}

Using Eqs. (\ref{eq:q1}) and (\ref{2ndform1}) we can solve Eq.(\ref{eq:secondordereq})
to obtain
\begin{multline}\label{eq:q2}
\hat{q}_2=-\frac{1}{\sigma-2i}\left[\frac{V^4A_2}{\rho_{0c}v_{Ac}^4|\Delta|x_0}\right.\\
\left.+\frac{V^8\left(p_e+A_1\right)^2\left(1+4\Omega_2\right)}{4\rho_{0c}^2v_{Ac}^8|\Delta|^2x_0^2(\sigma-i)^2}\right],
\end{multline}
where $\Omega_2=1/(\sigma-i)$
is the additional factor due to the nonlinear dispersion (as are all
subsequent values of $\Omega_i,\mbox{ }i>2$). We substitute the expression for
$\hat{q}_2$ into Eq. (\ref{eq:ujump1}) to find
\begin{equation}\label{eq:jumpinu22}
[\hat{u}_2]=-\frac{2\pi kV^5A_2}{\rho_{0c}v_{Ac}^4|\Delta|\cos^2\alpha},
\end{equation}
where the terms of the order of $k^2x_0^2$ are not indicated. To calculate
$A_2$ we compare the jump in the normal component of velocity across the dissipative layer
defined by Eqs. (\ref{eq:jumpinu21}) and (\ref{eq:jumpinu22}). This leads to $A_2=\mathscr{O}(k^2x_0^2)$.
This result implies that all quantities in the second order
approximation are zero outside the dissipative layer up to an
accuracy of $\mathscr{O}(kx_0)$. With this restriction the outgoing
wave remains monochromatic in the second order approximation. This
result coincides with the results of Ruderman \emph{et
al.}\citet{ruderman1997b}, Ballai \emph{et al.}\citet{ballai1998c},
Erd\'{e}lyi \emph{et al.}\citet{erdelyi2001} and Ruderman\citet{ruderman2000}
(this is especially surprising because in this paper nonlinearity is strong).

\subsection{Third order approximation}

The third order approximation with respect to $\zeta$
is governed by
\begin{multline}\label{eq:thirdordereq}
\sigma\frac{\da q_3}{\da\theta}-k^{-1}\frac{\da^2q_3}{\da\theta^2}
=-\frac{V^4}{\rho_{0c}v_{Ac}^4|\Delta|x_0}\frac{dP_{3c}}{d\theta}\\
-\frac{\da \left(q_1q_2\right)}{\da\theta}+\frac{\da q_1}{\da\sigma}\frac{\da q_2}{\da\theta}
+\frac{\da q_2}{\da\sigma}\frac{\da q_1}{\da\theta}.
\end{multline}
Taking into account the form of the solutions in the previous two orders of approximation
we can rewrite the second term on the right-hand side of Eq. (\ref{eq:thirdordereq})
as
\begin{equation}\label{eq:3rdform1}
\frac{\da\left(q_1q_2\right)}{\da\theta}=\frac{k}{2}\Re\left(
3i\hat{q}_1\hat{q}_2e^{3ik\theta}+i\hat{q}_1^{*}\hat{q}_2e^{ik\theta}\right),
\end{equation}
where $q_n=\Re\left(\hat{q}_ne^{ink\theta}+\hat{q}_n^{*}e^{-ink\theta}\right)$
and the asterisk denotes a complex conjugate. This result inspires us
to seek solutions in the third order approximation in the form
$g_3=\Re\left(\hat{g}_{31}e^{ik\theta}+\hat{g}_{33}e^{3ik\theta}\right)$,
where $g_3$ represents $P_3$, $u_3$ and $q_3$. Considering the
length of this paper we only calculate the $\hat{g}_{31}$
quantities, as it can be shown that $A_{33}=\mathscr{O}(k^2x_0^2)$.

In a similar manner as the first and second order approximations, we
find that the jump in the normal component of velocity across the
slow dissipative layer to be
\begin{multline}\label{eq:jumpinu31}
\left[\hat{u}_{31}\right]=\frac{i\kappa_iVA_{31}}{\rho_i\left(V^2-v_{Ai}^2\cos^2\alpha\right)}
+\frac{\kappa_eVA_{31}}{\rho_e\left(V^2-v_{Ae}^2\cos^2\alpha\right)}\\
-ikVA_{31}\mathscr{P}\int_{0}^{x_0}F^{-1}(x)\mbox{ }dx\\
-ikV(kx_0)\mathscr{P}\int_{0}^{x_0}\frac{\hat{h}_{31}(x)}{F(x)}\mbox{ }dx,
\end{multline}

To find $\hat{q}_{31}$ we must exploit Eqs. (\ref{eq:q1}),
(\ref{eq:q2}) and (\ref{eq:3rdform1}) to solve Eq.
(\ref{eq:thirdordereq}). The calculation is analogous to the first
and second order approximation calculations and we arrive at the
solution
\begin{multline}\label{eq:q31}
\hat{q}_{31}=-\frac{V^4A_{31}}{\rho_{0c}v_{Ac}^4|\Delta|x_0\left(\sigma-i\right)}\\
-\frac{V^{12}\left(p_e+A_1\right)|p_e+A_1|^2\left(1+2\Omega_{31}\right)}
{8\rho_{0c}^3v_{Ac}^{12}|\Delta|^3x_0^3\left(\sigma-i\right)^2\left(\sigma-2i\right)\left(\sigma^2+1\right)},
\end{multline}
where $\Omega_{31}$ and is given by
\begin{equation}
\Omega_{31}=\Omega_2\frac{\sigma^3-(8+7i)\sigma^2-(11+12i)\sigma-(44-5i)}{\left(\sigma-2i\right)\left(\sigma^2+1\right)},\nonumber
\end{equation}
We substitute this expression for $\hat{q}_{31}$ to find a second
definition for the jump in the normal component of velocity across
the slow dissipative layer (up to an accuracy of $kx_0$)
\begin{multline}\label{eq:2jumpinu31}
\left[\hat{u}_{31}\right]=-\frac{\pi kV^5A_{31}}{\rho_{0c}v_{Ac}^4|\Delta|\cos^2\alpha}\\
+\frac{\pi kV^{13}\left(p_e+A_1\right)|p_e+A_1|^2\left(27-8i\right)}{96\rho_{0c}^3v_{Ac}^{12}|\Delta|^3x_0^2\cos^2\alpha},
\end{multline}
Similar to the first two orders of approximation, we can compare
Eqs. (\ref{eq:jumpinu31}) with (\ref{eq:2jumpinu31}) to find the
coefficients $A_{31}$
\begin{equation}\label{eq:A3}
A_{31}=\frac{p_e^3\tau^3\mu^3\left(27-8i\right)\cos^4\alpha}
{12\pi^2V^2k^2x_0^2\left(\mu+i\upsilon\right)^2\left(\mu^2+\upsilon^2\right)},
\end{equation}
When calculating $A_{31}$ we have used the estimates
$\tau=\mathscr{O}(kx_0)$ and $kV\mathscr{P}\int_{0}^{x_0}F^{-1}(x)\mbox{ }dx=\mathscr{O}(kx_0)$,
and retain only the terms of lowest order with respect to $kx_0$, as
we have assumed that $kx_0\ll1$. Equation (\ref{eq:A3}) illustrates
that with an accuracy of up to $\mathscr{O}(kx_0)$ the outgoing
(reflected) wave remains monochromatic in the third order
approximation. Nevertheless, there is a slight alteration to the
amplitude of the fundamental harmonic of the outgoing wave from
$A_1$ to $A_1+\zeta^2A_{31}$. These results coincide, qualitatively,
with the findings by Ruderman \emph{et al.}\citet{ruderman1997b},
Ballai \emph{et al.}\citet{ballai1998c} and Erd\'{e}lyi \emph{et
al.}\citet{erdelyi2001}, however, $A_{31}$ is quantitatively larger than that
of previous studies and has an imaginary component. This implies that the amplitude of the wave is greater and
the phase of the correction is changed when compared with those studies. The expression for $A_{31}$, Eq.
(\ref{eq:A3}), is different to the ones they obtained because of the inclusion of dispersion through the Hall current.

\subsection{Higher order approximations}

In the fourth order of approximation the outgoing (reflected) wave becomes non-monochromatic.
This means the energy from this order of approximation no longer contribute to the fundamental harmonic, but to a higher one.
For full details of the calculation please refer to the Appendix.

Continuing calculations to even higher order approximations it can be shown
that the higher order harmonics (third, fourth, etc.) are generated in the outgoing (reflected)
fast wave. The pressure perturbation of the outgoing wave can be written as
\begin{equation}\label{eq:presspert}
P'=\epsilon\Re\left\{\sum_{n=1}^{\infty}\overline{A}_ne^{ink(\theta-\kappa_ex)}\right\}.
\end{equation}
The second harmonic only appears in the outgoing wave in the fourth order
approximation, whereas, higher harmonics appear in higher orders of approximation.
This implies that the estimate
$\overline{A}_n=\mathscr{O}(\zeta^3),\mbox{ }n\geq 2$ is valid.

\section{Solution inside the Alfv\'{e}n dissipative layer}

We can find the jump in the normal component of velocity at the
Alfv\'{e}n resonance explicitly, however, in an attempt to follow
the procedure in the last section (and to verify the theory), we
proceed to use the implicit form of the jump conditions. As the
governing equation (\ref{eq:governingalf}) is linear we only need to
calculate one order of approximation.

Although the Alfv\'{e}n resonant position is at $x=x_a$, compared
with $x=x_c$ for the slow resonant position, we can use some of
the same formulae as in the previous section. First, we look
for a solution in the form of $g_1=\Re(\hat{g}_1e^{ik\theta})$. In
Region I, we use Eqs. (\ref{eq:pressureeq1}) and (\ref{eq:ueq1}) to
represent the pressure and normal component of velocity
perturbations, respectively. For Region II, due to the first
connection formula, $\left[P\right]=0$, we can write the pressure perturbation as Eq.
(\ref{eq:pressureeq2}). We also find that Eqs.
(\ref{eq:pressureeq3}) and (\ref{eq:ueq3}) can be used to represent
the pressure and normal component of velocity perturbations,
respectively, in Region III. The fact we can employ the same
equations (as in slow resonance) in the three regions leads to one
of the definitions of the jump in the normal component of velocity
over the Alfv\'{e}n dissipative layer being defined as Eq.
(\ref{eq:jumpinu11}). It should come as no surprise that this
definition of the jump across the Alfv\'{e}n dissipative layer
coincides with the jump across the slow dissipative layer in the
first order approximation. We are using linear theory to obtain both expressions
and are not looking inside the, respective, dissipative layers', so
the forms should be identical.

To find $\hat{q}_a$, so that we find the other definition of the
jump in $u_a$, requires a different approach to the one utilized in
the section before. After Fourier analyzing Eq.
(\ref{eq:governingalf}), we are left with
\begin{equation}\label{eq:governingalf1}
i\sigma\hat{q}_a-\frac{d^2\hat{q}_a}{d\sigma_a^2}=\frac{ik\sin\alpha}{\rho_{0a}|\Delta_a|}P_a.
\end{equation}
To solve Eq. (\ref{eq:governingalf1}) we introduce the Fourier
transform with respect to $\sigma$:
\begin{equation}\label{eq:fouriertransformsigma}
\mathscr{F}\left[f(\sigma)\right]=\int_{-\infty}^{\infty}f(\sigma)e^{-i\sigma
r}\mbox{ }d\sigma.
\end{equation}
Then from Eq. (\ref{eq:governingalf1}) we have
\begin{equation}\label{eq:governingalf2}
\frac{d\mathscr{F}[\hat{q_a}]}{dr}-r^2\mathscr{F}[\hat{q_a}]=-\frac{2\pi
ik\sin\alpha\left(p_e+A\right)}{\rho_{0a}|\Delta_a|}\delta(r),
\end{equation}
where $\delta(r)$ is the delta-function. We find that the solution
to Eq. (\ref{eq:governingalf2}) that is bounded for
$|r|\rightarrow\infty$ is
\begin{equation}\label{eq:qa}
\mathscr{F}[\hat{q_a}]=\frac{2i\pi
k\sin\alpha(p_e+A)}{\rho_{0a}|\Delta_a|}H(-r)e^{r^3/3}.
\end{equation}
Here $H(r)$ denotes the Heavyside function. It was shown by Ruderman
and Goossens\citet{ruderman1997b} that
\begin{equation}\label{eq:cauchyrule}
\mathscr{P}\int_{-\infty}^{\infty}f(\sigma)\mbox{
}d\sigma=\frac{1}{2}\left(\lim_{r\rightarrow
+0}\mathscr{F}[f]+\lim_{r\rightarrow -0}\mathscr{F}[f]\right).
\end{equation}
With the aid of Eqs. (\ref{eq:ujumpalf}), (\ref{eq:qa}) and
(\ref{eq:cauchyrule}) we find that
\begin{equation}\label{eq:ua1}
[\hat{u}_a]=-\frac{\pi kV(p_e+A)\sin^2\alpha}{\rho_{0a}|\Delta_a|}.
\end{equation}

Comparing Eqs. (\ref{eq:jumpinu11}) and (\ref{eq:ua1}) we derive
that
\begin{equation}\label{eq:A}
A=-p_e\frac{\tau_a-\mu+i\upsilon}{\tau_a+\mu+i\upsilon}+\mathscr{O}(k^2x_0^2),
\end{equation}
where
$\tau_a=\pi kV\sin^2\alpha/(\rho_{0a}|\Delta_a|)$,
and $\mu$ and $\upsilon$ have their forms given by Eq. (\ref{eq:taumuupsilon}).
However, their values are different for the two resonances.

\section{Coefficient of Wave Absorption}

The coefficient of wave absorption is defined as
$\Gamma=(\Pi_{\rm{in}}-\Pi_{\rm{out}})/\Pi_{\rm{in}}$,
where $\Pi_{\rm{in}}$ and $\Pi_{\rm{out}}$ are the normal
components of the energy fluxes, averaged over a period,
of the incoming and outgoing waves, respectively. It is
straightforward to obtain that
\begin{equation}\label{eq:Gamma}
\Gamma=1-\frac{1}{p_e^2}\sum_{n=1}^{\infty}|\overline{A}_n|^2
\approx\Gamma_{\rm{L}}+\zeta^2\Gamma_{\rm{ND}},
\end{equation}
where $\Gamma_{\rm{L}}$ is the linear coefficient of wave absorption
and $\Gamma_{\rm{ND}}$ is the nonlinear and dispersive correction.
Note that $\Gamma_{\rm{ND}}$ is multiplied by the small factor
$\zeta^2$ which means that this term will provide small corrections
to linear results.

Carrying out calculations we find at the slow resonance, in
agreement with linear theory, that
\begin{equation}
\Gamma_{\rm{L}}=\frac{4\tau\mu}{\mu^2+\upsilon^2}+\mathscr{O}(k^2x_0^2).
\end{equation}
The coefficient $\Gamma_{\rm{ND}}$ is defined as
$\Gamma_{\rm{ND}}=-(2/p_e^2)\Re\left\{A_1^{*}A_{31}\right\}$,
which can be rewritten using Eqs. (\ref{eq:A1}) and (\ref{eq:A3}) as
\begin{equation}
\Gamma_{\rm{ND}}=-\frac{27p_e^2\tau^3\mu^3\cos^4\alpha}{6\pi^2V^2k^2x_0^2\left(\mu^2+\upsilon^2\right)^2}
+\mathscr{O}(k^2x_0^2).
\end{equation}
Both $\Gamma_{\rm{L}}$ and $\Gamma_{\rm{ND}}$ are of the order of
$kx_0$. This result is qualitatively the same as Ruderman {et
al.}\citet{ruderman1997b} and Ballai \emph{et
al.}\citet{ballai1998c} results, however, the nonlinear correction
is different. In fact, it is $270\%$ times larger due to the Hall current having a dominant effect around the resonance.
Moreover, the dispersion in the slow dissipative layer
causes a further reduction in the coefficient of energy absorption,
in comparison to the nonlinear regime alone.

At the Alfv\'{e}n resonance dynamics can be described within the linear framework. Hence, using Eqs.
(\ref{eq:A}) and (\ref{eq:Gamma}) we obtain that
\begin{equation}\label{eq:gammaalf}
\Gamma_a=\frac{4\tau_a\mu}{(\tau_a+\mu)^2+\upsilon^2}.
\end{equation}
Numerical verification of these results requires much more work than would first appear, and
as such our next paper is to concentrates on this and further numerical analysis.

\section{Conclusions}

In the present paper we have investigated (i) the effect of
nonlinearity and dispersion on the interaction of fast
magnetoacoustic (FMA) waves with a one-dimensional inhomogeneous
magnetized plasma with strongly anisotropic transport processes in
the slow dissipative layer (ii) the interaction of FMA waves with
Alfv\'{e}n dissipative layers. The study is based on the nonlinear
theory of slow resonance in strongly anisotropic and
dispersive plasmas developed by Clack and Ballai\citet{clack2008}
and the theory of Alfv\'{e}n resonance developed by
Clack \emph{et al.}\citet{clack2008b}.

We have assumed that (i) the thickness of the slab containing the
inhomogeneous plasma (Region II) is small in comparison with the
wavelength of the incoming fast wave (i.e. $kx_0\ll1$); and (ii) the
nonlinearity in the dissipative layer is weak - the nonlinear term
in the equation describing the plasma motion in the slow dissipative
layer can be considered as a perturbation and nonlinearity gives
only a correction to the linear results.

Applying a regular perturbation method, analytical solutions in the
slow dissipative layer are obtained in the form of power expansions
with respect to the nonlinearity parameter $\zeta$. Our main results
are the following: Nonlinearity in the dissipative layer generates
higher harmonic contributions to the outgoing (reflected) wave in
addition to the fundamental one. The dispersion does not alter this,
however, the phase and amplitude of some of the higher harmonics are
different from the standard nonlinear counterpart (see discussions before).
Dispersion in the dissipative
layer further decreases the coefficient of the wave energy
absorption. The factor of alteration to the \emph{nonlinear}
correction of the coefficient of wave absorption due to dispersion
is $270\%$. Remember, however, that the nonlinear correction is
multiplied by the small parameter $\zeta^2$, so the effect to the
overall coefficient of wave energy absorption is still small.

Calculating the coefficient of wave absorption at the Alfv\'{e}n
resonance confirms the linear theory of the past and verifies the
approach taken to be correct. As our physical set-up of the problem
(for the Alfv\'{e}n resonance) matches the typical conditions found
in the solar corona, these results can be applied to it. The
equilibrium state of the problem (for the slow resonance) can match
conditions found in the upper chromosphere, where FMA waves may
interact with slow dissipative layers, and if the reduction in the
coefficient of wave energy absorption persists to the strong
nonlinear case (as with the long wavelength approximation found by
Ruderman \citep{ruderman2000}) dispersion may have further
implications to the resonant absorption in the solar atmosphere.

In a forthcoming paper, we shall theoretically and numerically investigate coupled
resonances, which builds from the work in the present paper to obtain a more realistic model
for a solar physical description. In the same paper we will
numerically analyze the absorption of fast waves at the Alfv\'{e}n
resonance as a possible scenario of the interaction of global fast waves
(modelling EIT waves) and coronal loops.

\section*{ACKNOWLEDGEMENTS}

The authors wish to thank M.~S. Ruderman for helpful comments and discussions.
C.~T.~M. Clack would like to thank STFC (Science and Technology Facilities
Council) for the financial support provided. I. Ballai acknowledges
the financial support by NFS Hungary (OTKA, K67746) and The National
University Research Council Romania (CNCSIS-PN-II/531/2007).

\section*{APPENDIX: DETAILS FOR CALCULATION OF FOURTH ORDER APPROXIMATION}

In the fourth order approximation Eq. (\ref{eq:governingproper}) gives
\begin{multline}\label{eq:fourthordereq}\tag{A1}
\sigma\frac{\da q_4}{\da\theta}-k^{-1}\frac{\da^2q_4}{\da\theta^2}
=-\frac{V^4}{\rho_{0c}v_{Ac}^4|\Delta|x_0}\frac{dP_{4c}}{d\theta}
+\frac{\da q_2}{\da\sigma}\frac{\da q_2}{\da\theta}\\
-\frac{\da}{\da\theta}\left(q_1q_3+\frac{1}{2}q_2^2\right)
+\frac{\da q_1}{\da\sigma}\frac{\da q_3}{\da\theta}
+\frac{\da q_3}{\da\sigma}\frac{\da q_1}{\da\theta}.
\end{multline}
We can rewrite the third term on the right-hand side of Eq.
(\ref{eq:fourthordereq}) using our knowledge about the first three
orders of approximation, so
\begin{multline}\label{eq:4thform1}\tag{A2}
\frac{\da}{\da\theta}\left(q_1q_3+\frac{1}{2}q_2^2\right)=
k\Re\left\{i\left(\hat{q}_{1}\hat{q}_{31}+\hat{q}_{1}^{*}\hat{q}_{33}\right)e^{2ik\theta}\right.\\
\left.+i\left(2\hat{q}_{1}\hat{q}_{33}+\hat{q}_{2}^2\right)e^{4ik\theta}\right\}.
\end{multline}
This equation contains terms proportional to $e^{2ik\theta}$ and
$e^{4ik\theta}$, so we can anticipate the solution to Eq.
(\ref{eq:fourthordereq}) to be of the form
$g_4=\Re\left(\hat{g}_{42}e^{2ik\theta}+\hat{g}_{44}e^{4ik\theta}\right)$,
where $g_4$ represents $P_4$, $u_4$ and $q_4$. We calculate
the fourth order approximation to demonstrate that nonlinearity and dispersion
in the dissipative layer generates overtones in the outgoing (reflected) fast wave.
For brevity, we shall only derive the terms proportional to $e^{2ik\theta}$, but for completeness
we note that it can be shown that terms proportional to $e^{4ik\theta}$ are only
present in the solution inside the dissipative layer.

Using the continuity conditions at $x=0$ and $x=x_0$ we find the jump in the
normal component of velocity across the dissipative layer to be
\begin{multline}\label{eq:jumpinu42}\tag{A4}
\left[\hat{u}_{42}\right]=\frac{i\kappa_iVA_{42}}{\rho_i\left(V^2-v_{Ai}^2\cos^2\alpha\right)}
+\frac{\kappa_eVA_{42}}{\rho_e\left(V^2-v_{Ae}^2\cos^2\alpha\right)}\\
-2ikVA_2\mathscr{P}\int_{0}^{x_0}F^{-1}(x)\mbox{ }dx\\
-2ikV(kx_0)\mathscr{P}\int_{0}^{x_0}\frac{\hat{h}_2(x)}{F(x)}\mbox{ }dx.
\end{multline}

It is straightforward, but longwinded, to derive $\hat{q}_{42}$, so
we skip all intermediate steps and give the result
\begin{multline}\label{eq:q42}\tag{A5}
\hat{q}_{42}=\frac{-1}{\sigma-2i}\left\{\frac{V^4A_{42}}{\rho_{0c}v_{Ac}^4|\Delta|x_0}
+\frac{V^8\left(p_e+A_1\right)A_{31}\left(1+\Omega_2\right)}{2\rho_{0c}v_{Ac}^8|\Delta|^2x_0^2(\sigma-i)^2}\right.\\
\left.+\frac{V^{16}\left(p_e+A_1\right)^2|p_e+A_1|^2\left(12-\Omega_{42}\right)}{96\rho_{0c}^4v_{Ac}^{16}|\Delta|^4x_0^4(\sigma-i)^3
(\sigma-3i)\left(\sigma^2+1\right)}\right\},
\end{multline}
with $\Omega_{42}=f(\sigma)$, where $f(\sigma)\rightarrow 0$ as
$\sigma\rightarrow\infty$, is the contribution due to the Hall
effect. As it is not essential for forthcoming calculations, its
exact form is not given here. The substitution of $\hat{q}_{42}$
into Eq. (\ref{eq:ujump1}) yields
\begin{multline}\label{eq:2jumpinu42}\tag{A6}
\left[\hat{u}_{42}\right]=-\frac{2\pi kV^5A_{42}}{\rho_{0c}v_{Ac}^4|\Delta|\cos^2\alpha}\\
+0.082\times\frac{\pi kV^{17}(p_e+A_1)^2|p_e+A_1|^2}{\rho_{0c}^4v_{Ac}^{16}|\Delta|^4x_0^3\cos^2\alpha},
\end{multline}
Comparing Eqs. (\ref{eq:jumpinu42}) and (\ref{eq:2jumpinu42}) we obtain that
\begin{equation}\label{eq:A42}\tag{A7}
A_{42}=1.279\times\frac{p_e^4\tau^4\mu^4\cos^6\alpha}{\pi^3V^3k^3x_0^3
\left(\mu+i\upsilon\right)^3\left(\mu^2+\upsilon^2\right)}.
\end{equation}
Here we have used the same estimations that were utilized for
calculating $A_{31}$ in the third order approximation and retain
only the largest order terms with respect to $kx_0$. It is clear
from this result that the outgoing wave becomes non-monochromatic in
the fourth order approximation. We can also observe that the
second harmonic appears in addition to the fundamental mode.

This result parallels the results obtained by Ruderman
\emph{et al.}\citet{ruderman1997b} and Ballai \emph{et
al.}\citet{ballai1998c}. However, Eq. (\ref{eq:A42}) shows that the
\emph{phase} is inverted and the \emph{amplitude} of the second harmonic is approximately $30$
times greater than theirs due to the presence of the Hall effect. Remember, though, that this amplitude
is multiplied by a very small term, $\zeta^3$, which means the overall correction is very small.

\end{document}